  \documentclass[final,5p,times,twocolumn]{elsarticle}

\usepackage{epsfig}

\usepackage{amssymb}
 \usepackage{amsthm}
 \usepackage{amsmath}
\usepackage{xcolor}  

\journal{}

\begin{document}

\begin{frontmatter}



\title{CROSS-SECTIONS FOR THE ${^{27}\!\rm{Al}}(\gamma,\textit{x})^{22}\rm{Na}$  MULTICHANNEL REACTION WITH THE 28.3 MeV DIFFERENCE OF THE REACTION THRESHOLDS}

\author{ O.S. Deiev} 

\author{I.S. Timchenko\corref{cor1}}
\ead{timchenko@kipt.kharkov.ua}
 \cortext[cor1]{Corresponding author}
\author{S.N. Olejnik, A.N. Vodin, \\ M.I. Ayzatskiy, V.A. Kushnir, V.V. Mitrochenko, S.A. Perezhogin}
\address{National Science Center "Kharkiv Institute of Physics and Technology", \\
 1 Akademicheskaya St., 61108 Kharkiv, Ukraine}

\begin{abstract}
The bremsstrahlung flux-averaged cross-sections $\langle{\sigma(E_{\rm{\gamma max}})}\rangle$ and the cross-sections per equivalent photon $\langle{\sigma(E_{\rm{\gamma max}})_{\rm{Q}}}\rangle$ were first measured for the photonuclear multichannel reaction ${^{27}\!\rm{Al}}(\gamma,\textit{x})^{22}\rm{Na}$  at end-point bremsstrahlung gamma energies ranging from 35~MeV to 95 MeV. The experiments were performed with the beam from the NSC KIPT electron linear accelerator LUE-40 with the use of the $\gamma$-activation technique. The bremsstrahlung quantum flux was calculated with the program GEANT4.9.2 and, in addition, was monitored by means of the $^{100}\rm{Mo}(\gamma,n)^{99}\rm{Mo}$ reaction. The flux-averaged cross-sections were calculated using the partial cross-section $\sigma(E)$ values computed with the TALYS1.95 code for different level density models. Consideration is given to special features of calculating the cross-sections $\langle{\sigma(E_{\rm{\gamma max}})}\rangle$ and  $\langle{\sigma(E_{\rm{\gamma max}})_{\rm{Q}}}\rangle$ for the case of the final nucleus $^{22}\rm{Na}$ production in the $^{27}\!\rm{Al}$ photodisintegration reaction via several partial channels $\textit{x}$: $\rm{n}\alpha + dt + npt + 2n{^{3}\rm{He}} + n2d + 2npd + 2p3n$. 
\end{abstract}




\end{frontmatter}

\section{Introduction}
\label{Int}

Nowadays, vigorous activities are taken on to investigate by experiment the nuclear photodisintegration in the energy range above the giant dipole resonance (GDR) and up to the pion production threshold ($E_{\rm{th}} \approx 145$~MeV) \cite{1,2,3,4,5,6,7}. The interest in this energy region is caused by the change in the mechanism of photon-nucleus interaction occurring there. That opens up a new line of research that would provide a detailed knowledge of the competition between two mechanisms of nuclear photodisintegration, viz., through GDR excitation and quasideuteron photoabsorption. 

The data obtained from the photonuclear experiment on bremsstrahlung beams are generally represented in terms of the relative yield or the integrated reaction cross-section  \cite{2}, the average cross-section  $\langle{\sigma(E_{\rm{\gamma max}})_{\rm{Q}}}\rangle$ \cite{3,4,5,6,7}, or the cross-section per equivalent photon  $\langle{\sigma(E_{\rm{\gamma max}})_{\rm{Q}}}\rangle$ \cite{4,8}. The $\langle{\sigma(E_{\rm{\gamma max}})}\rangle$ value reflects the energy dependence of the cross-section $\sigma(E)$ for the reaction under study, and is also related to the gamma flux in the range from the reaction threshold  $E_{\rm{th}}$  to the end-point bremsstrahlung energy $E_{\rm{\gamma max}}$. The use of $\langle{\sigma(E_{\rm{\gamma max}})}\rangle$ value provides a more detailed consideration of the cross-section behavior against the bremsstrahlung gamma energy, because $\langle{\sigma(E_{\rm{\gamma max}})}\rangle$ is insensitive to the low-energy part of the bremsstrahlung spectrum. At the same time, the cross-section per equivalent quantum $\langle{\sigma(E_{\rm{\gamma max}})_{\rm{Q}}}\rangle$  is proportional to the reaction yield, and always increases with the $E_{\rm{\gamma max}}$ increase. Among the disadvantages of the use of these cross-sections, one can mention the effect of bremsstrahlung gamma quanta of energies below the threshold of the reaction under study. 

This becomes of importance in the interpretation of the data gained from the experiments with application of the induced activity method, where the gamma-radiation of the reaction product nucleus is registered. Thus in photoneutron reactions, the final nucleus is determined by the quantity of knocked-on neutrons and is formed at the energy equal to/higher than the threshold energy $E_{\rm{th}}$ of the reaction under study, which is constant. In the case of multiparticle photonuclear reactions with the charged-particle presence in the exit channel, the reaction product can be produced via several partial channels, each of which has its own reaction threshold value. For example, in the $^{27}\!\rm{Al}(\gamma,\textit{x})^{24}\rm{Na}$ reaction the $^{24}\rm{Na}$ nucleus is formed via three reaction channels, the difference of their threshold energies $E_{\rm{th}}$ being 7.7~MeV. It has been demonstrated in ref.~\cite{4} that in the calculation of the total flux-averaged cross-section $\langle{\sigma(E_{\rm{\gamma max}})}\rangle$ for the given reaction it is necessary to take into account the threshold value of each reaction channel. The attempt of using in the calculation only the $E_{\rm{th}}$ value of the dominant reaction channel leads to an essential error in the $\langle{\sigma(E_{\rm{\gamma max}})}\rangle$ value in the neighborhood of the minimal threshold of the reaction studied. The application of the $\langle{\sigma(E_{\rm{\gamma max}})_{\rm{Q}}}\rangle$ value for representing the data of this reaction appears to be more convenient \cite{4}.

   The situation becomes still more complicated for the $^{27}\!\rm{Al}(\gamma,\textit{x})^{22}\rm{Na}$ reaction, where the final product nucleus can simultaneously be formed in seven partial channels: $\textit{x} = \rm{n}\alpha + dt + npt + 2n{^{3}\rm{He}} + n2d + 2npd + 2p3n$, with difference of their threshold energies reaching 28.3~MeV. This significant difference in the threshold energies of the reaction channels makes it possible to investigate the contribution of different channels to the production of the final product nucleus $^{22}\rm{Na}$. For example, at energies between 22.5 and 60 MeV it is the channel of the $^{27}\!\rm{Al}(\gamma,\rm{n}\alpha)^{22}\rm{Na}$ reaction that is dominant, whereas at energies above 70 MeV a considerable contribution also comes from the $^{27}\!\rm{Al}(\gamma,2p3n)^{22}\rm{Na}$ reaction. 

    Previously, the photodisintegration of the $^{27}\!\rm{Al}$ nucleus with production of $^{22}\rm{Na}$ has been investigated in works \cite{8,9}. Those authors have determined the experimental bremsstrahlung flux-averaged cross-section values per equivalent photon, $\langle{\sigma(E_{\rm{\gamma max}})_{\rm{Q}}}\rangle$, at end-point bremsstrahlung gamma-quantum energies ranging from 250 to 1000~MeV.  

The present work has been concerned with measuring the average cross-sections  $\langle{\sigma(E_{\rm{\gamma max}})}\rangle$ and  $\langle{\sigma(E_{\rm{\gamma max}})_{\rm{Q}}}\rangle$ for the multichannel photonuclear reaction $^{27}\!\rm{Al}(\gamma,\textit{x})^{22}\rm{Na}$ in the energy range $E_{\rm{\gamma max}} = 35$--95~MeV. Consideration has been given to some special features of the calculation of $\langle{\sigma(E_{\rm{\gamma max}})}\rangle$ and  $\langle{\sigma(E_{\rm{\gamma max}})_{\rm{Q}}}\rangle$ values. Comparison has been made with the calculations based on the data for the partial cross-sections $\sigma(E)$ of the reaction channels ($\rm{n}\alpha + dt + npt + 2n{^{3}\rm{He}} + n2d + 2npd + 2p3n$), computed with the TALYS1.95 code \cite{10} for different level density models.

\section{Experimental procedure}
\label{Exp proced}

\begin{figure}[h]
   	\center{\includegraphics[scale=0.3]{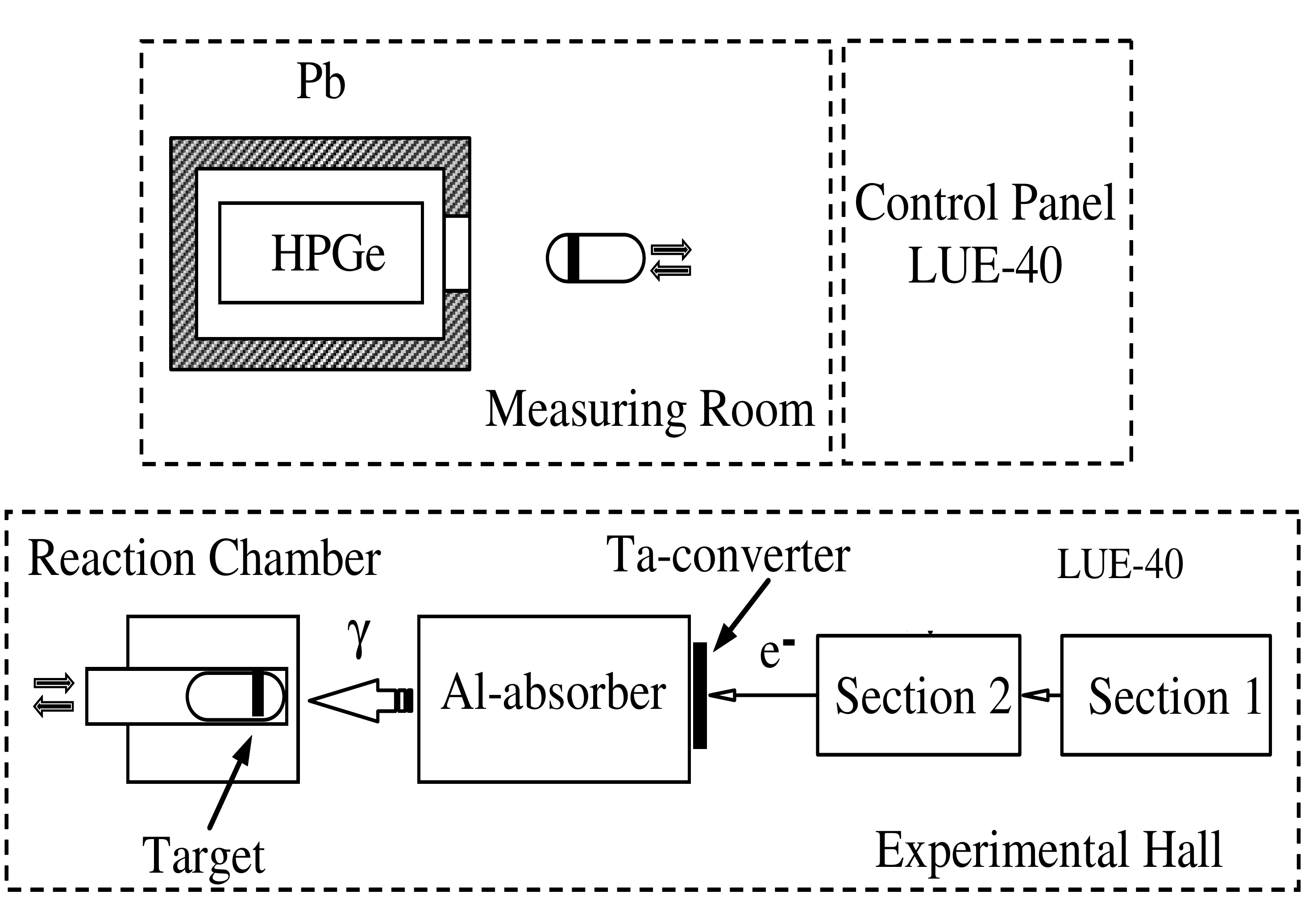}}
   	\caption{Experimental schematic diagram including three units framed with a dashed line.  Above -- the measuring room and the control panel of the LUE-40 accelerator, below -- the experimental hall.}
   	\label{fig1}
   \end{figure}

The experiments were performed on the bremsstrahlung $\gamma$-beam from the electron linear accelerator LUE-40 NSC KIPT with the use of the method of induced activity of the final product nucleus of the reaction, similar to that of refs. \cite{3,11}. The schematic diagram of the experiment is presented in Fig.~\ref{fig1}.
   
    The accelerator parameters enable the change in the energies of accelerated electrons in the range of $E_e = $ 30 to 100 MeV at the average beam current $I_e = 3~\mu$A. With that, the electron energy spectrum width (FWHM) makes $\triangle E_e/E_e \sim 1$--$1.5 \%$ at pulse repetition frequency of 50~Hz and a pulse length of $10~\mu$s \cite{U1,U2}.

   The bremsstrahlung flux was generated as the pulsed electron beam passed through a 1.05~mm thick tantalum metal plate (radiation length of Ta being $\sim\!4.1$~mm). The Ta-converter was fixed on aluminum cylinder, 100~mm in diameter and 150~mm in thickness. The aluminum cylinder was used for absorption of the electrons that have passed through the converter. The application of the Al-absorber resulted in bremsstrahlung spectrum distortion and additional generation of neutrons.
  
    The bremsstrahlung flux was computed with the GEANT4.9.2 code \cite{12}. The program code GEANT4.9.2, PhysList \textit{G4LowEnergy} allows one to perform calculations with correct consideration of all physical processes for the case of an amorphous target. Similarly, GEANT4.9.2, PhysList \textit{QGSP\_BIC\_HP} makes it possible to calculate the neutron yield due to photonuclear reactions from targets of different thickness and atomic charge.

For the experiments, $^{\rm{nat}}$Mo and $^{27}\!\rm{Al}$ targets were prepared in the form of thin discs of the like diameters 8~mm and the thicknesses of 0.1~mm for molybdenum and 1~mm for aluminum, that corresponding to the masses $m \approx$ 60 and 136 mg, respectively. The both targets were simultaneously exposed to bremsstrahlung gamma quanta for the time $t_{\rm{irr}}$ = 30~min at all electron energy values. The targets were delivered to the reaction chamber in a special aluminum capsule by means of the pneumatic conveyor system. After the exposure, the both targets were transferred to the measuring room, where they were removed from the capsule, and the induced $\gamma$-activity spectra of the targets were registered one-by-one by means of the HPGe detector. The measurement time was $t_{\rm{meas}}$ = 30~min for the Mo target, and $t_{\rm{meas}}$ = 1 to 4 days for the aluminum target. The induced activity of the aluminum targets was measured after cooling the targets for 5 to 50 days, thereby eliminating the contribution of intense spectrum lines of the $^{24}\rm{Na}$ nucleus ($E_{\gamma} = 1368.6$ and 2754.0~keV, $T_{1/2}$ = $14.96$~h).

\begin{figure*}[h]
	\center{\includegraphics[scale=0.45]{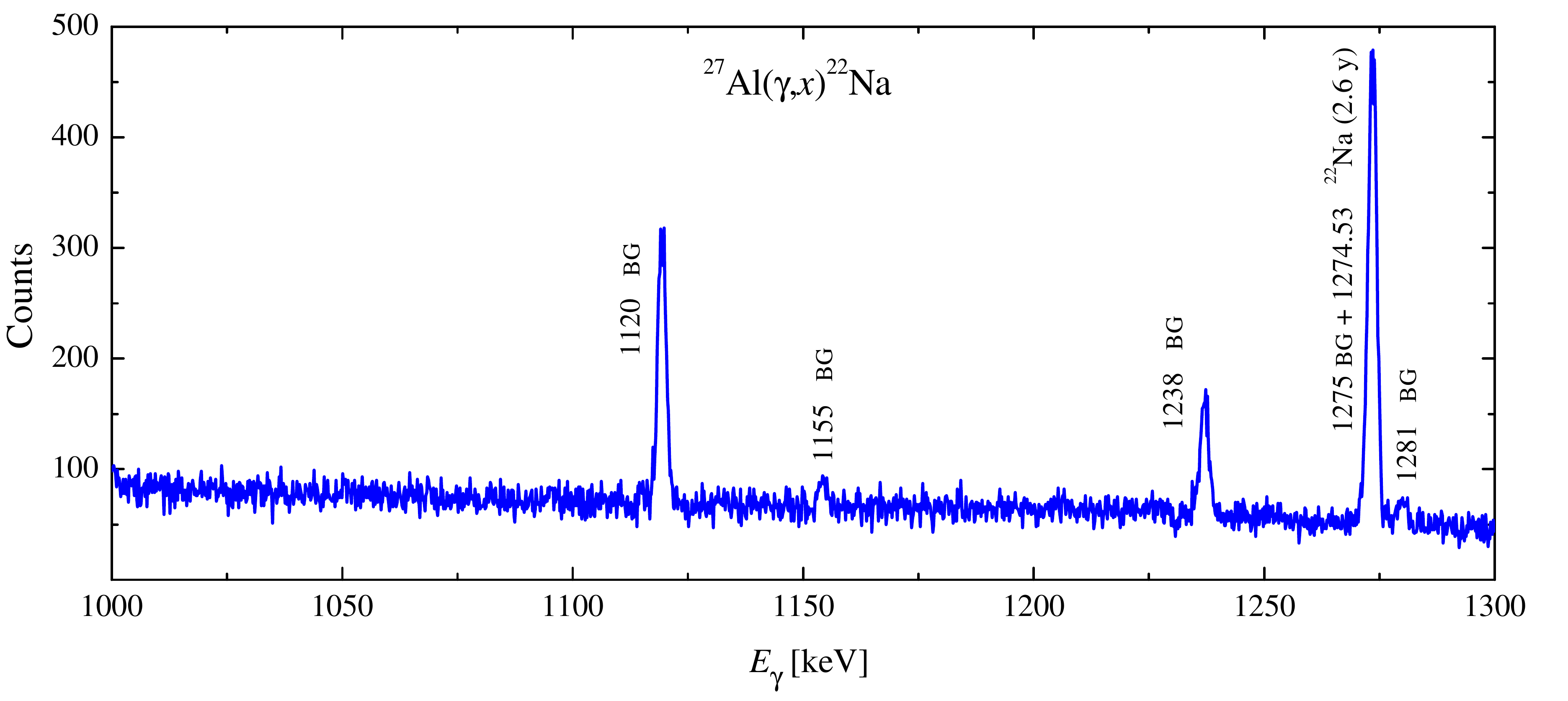}}
	\caption{Part of gamma-ray spectrum from the activated $^{27}\!\rm{Al}$ target of mass 135.486~mg, after exposure to $\gamma$-quanta flux with the end-point energy of the bremsstrahlung spectrum $E_{\rm{\gamma max}} = 85.6$~MeV, $t_{\rm{cool}}$ = 30 days and $t_{\rm{meas}}$ = 85~h. The spectrum fragment ranges from 1000 to 1300~keV. The background $\gamma$-lines peaks  are indicated by the letters BG.}
	\label{fig2}
\end{figure*}

   The resolution of the HPGe detector Canberra GS-2018 was 1.8~keV (FWHM) for the $E_{\gamma} = 1332.5$~keV $\gamma$-line of $^{60}$Co, and its efficiency was 20\% relative to the NaI(Tl) detector, 300~mm in diameter and 300~mm in thickness. The standard radiation sources $^{22}$Na, $^{60}$Co, $^{133}$Ba, $^{137}$Cs, $^{152}$Eu and $^{241}\!\rm{Am}$ were used for energy/efficiency calibration of the spectrometry channel. In $\gamma$-ray spectrum measurements, the dead time of the spectrometry channel was no more than 3 to 5\%; it was determined through choosing the appropriate distance between the irradiated sample and the HPGe detector. The $\gamma$-ray spectra were analyzed using the Canberra GENIUS2000 software \cite{13}. Figure~\ref{fig2} shows a typical fragment of the $\gamma$-radiation spectrum from the aluminum target.
      
   The $^{22}$Na $\gamma$-activity studies of the $^{27}\!\rm{Al}(\gamma,\textit{x})^{22}\rm{Na}$ reaction were performed through the use of the 
$E_{\gamma} = 1274.53$~keV $\gamma$-line. The characteristics of this $\gamma$-line are presented in Table~\ref{tab1}. The production of $^{22}$Na is possible in seven partial channels of the reaction: $\textit{x} = \rm{n}\alpha + dt + npt + 2n{^{3}\rm{He}} + n2d + 2npd + 2p3n$. The threshold energy difference between the partial channels of the $^{27}\!\rm{Al}(\gamma,\textit{x})^{22}\rm{Na}$ reaction reaches 28.3 MeV, this being rather significant in estimating the bremsstrahlung gamma flux value, which is used in the computation of the total average cross-section $\langle{\sigma(E_{\rm{\gamma max}})}\rangle$ value. 

 \begin{table*}[h]
      	\caption{\label{tab1} Spectroscopic data from ref.~\cite{14} for the product-nuclei from the reactions $^{27}\!\rm{Al}(\gamma,\textit{x})^{22}\rm{Na}$ and $^{100}\rm{Mo}(\gamma,n)^{99}\rm{Mo}$  }
      	\centering
      	\begin{tabular}{ccccc}
      		\hline	\vspace{1ex}
      		Nuclear reaction & $E_{\rm{th}}$,~MeV & $T_{1/2}$ & $E_{\gamma}$,~keV & $I_{\gamma}$, \% \\ 	\hline
      	  \vspace{1ex}	   
      \begin{tabular}{c} $^{27}\!\rm{Al}(\gamma,n\alpha)^{22}\rm{Na}$ \\ $^{27}\!\rm{Al}(\gamma,dt)^{22}\rm{Na}$ \\ $^{27}\!\rm{Al}(\gamma,npt)^{22}\rm{Na}$ \\ $^{27}\!\rm{Al}(\gamma,2n{^3\rm{He}})^{22}\rm{Na}$\\ $^{27}\!\rm{Al}(\gamma,n2d)^{22}\rm{Na}$ \\ $^{27}\!\rm{Al}(\gamma,2npd)^{22}\rm{Na}$  \\ $^{27}\!\rm{Al}(\gamma,2p3n)^{22}\rm{Na}$
      \end{tabular}& 
  \begin{tabular}{c} 22.51\\ 40.10 \\ 42.33 \\ 43.09 \\ 46.36 \\ 48.58 \\ 50.81
  \end{tabular}&       	  $2.6019 \pm 0.0004$~y & 
      	 \begin{tabular}{c}1274.53 
      	 \end{tabular}& 
           	 \begin{tabular}{c}$99.944 \pm 0.014$
     \end{tabular}\\   
 \hline
$^{100}\rm{Mo}(\gamma,n)^{99}\rm{Mo}$ & 8.29 & $65.94 \pm 0.01$~h & 
      		\begin{tabular}{c}739.50 
      		\end{tabular}& 
      	\begin{tabular}{c}$12.13 \pm 0.12$ 
      	\end{tabular} \\	
         \hline
      	\end{tabular}	        
  \end{table*}

The bremsstrahlung $\gamma$-flux monitoring by the $^{100}\rm{Mo}(\gamma,n)^{99}\rm{Mo}$ reaction yield was performed by comparing the experimentally obtained average cross-section values with the computation data. To determine the experimental $\langle{\sigma(E_{\rm{\gamma max}})}\rangle_{\rm{exp}}$ values, we have used the activity $\triangle A$ for the $\gamma$-line of energy $E_{\gamma} = 739.50$~keV, and the absolute intensity $I_{\gamma}$ = 12.13\% (see Table 1). The average cross-section $\langle{\sigma(E_{\rm{\gamma max}})}\rangle_{\rm{th}}$ values were computed with the cross-sections $\sigma(E)$ from the TALYS1.95 code. The obtained normalization factor $k = \langle{\sigma(E_{\rm{\gamma max}})}\rangle_{\rm{th}} / \langle{\sigma(E_{\rm{\gamma max}})}\rangle_{\rm{exp}}$ represents the deviation of the GEANT4.9.2-computed bremsstrahlung $\gamma$-flux from the real $\gamma$-flux incident on the target. The thus found $k$ values, which varied within 1.08--1.15, were used for normalization of the cross-sections for the $^{27}\!\rm{Al}$ nucleus. 

The Ta-converter and Al-absorber, used in the experiment, generate the neutrons that can cause the reaction $^{100}\rm{Mo}(n,2n)^{99}\rm{Mo}$. Calculations were made of the neutron energy spectrum and the fraction of neutrons of energies above the threshold of this reaction, similarly to \cite{15}. The contribution of the $^{100}\rm{Mo}(n,2n)^{99}\rm{Mo}$ reaction to the induced activity of the $^{99}\rm{Mo}$ nucleus has been estimated and has been found to be negligible as compared to the contribution of $^{100}\rm{Mo}(\gamma,n)^{99}\rm{Mo}$. The contribution of the reaction $^{100}\rm{Mo}(\gamma,p)^{99}\rm{Nb}$, $^{99}\rm{Nb} \xrightarrow{\beta^-}$$^{99}\rm{Mo}$ is also negligible.

   The difficulties in measurements of the  $^{27}\!\rm{Al}(\gamma,\textit{x})^{22}\rm{Na}$ reaction are partially due to a long half-life of the $^{22}\rm{Na}$ nucleus ($T_{1/2}$ = 2.6 years), that leads to the necessity of measuring the target activity for a long time. Under such conditions, consideration must be given to the contribution of the natural background lines to the spectrum of the induced activity spectrum of the sample. As indicated in ref.~\cite{16}, the background spectrum exhibits a low-intensity line at energy of $\sim$1275 keV, the contribution of which to $\triangle A$  of the 1274.53 keV $\gamma$-line under study becomes essential at long-time measurement. In our experiment, this occurred with the spectra from the targets irradiated at low end-point bremsstrahlung gamma energies $E_{\rm{\gamma max}}$ $\sim$ 35--45 MeV. In this case, the percentage of background counts ranged up to 20--40\% with the result that the statistical error accordingly increased. 

   The accuracy of average cross-sections measurements was determined as a quadratic sum of statistical and systematic errors. The statistical error of the observed  $\gamma$-line is mainly due to the statistical calculation, and is estimated to vary within 3\% to 20\%.

 The systematic errors are associated with the uncertainties of: 1) the exposure time, 0.25\%; 2) the electron current, 0.5\%; 3) the $\gamma$-radiation detection efficiency, $\sim$  2 to 2.5\%, which is mainly attributed to the uncertainties of gamma radiation sources and data approximation; 4) normalization of experimental data to the yield of the monitoring reaction $^{100}\rm{Mo}(\gamma,n)^{99}\rm{Mo}$, $\sim$  2--2.5\%. 

It should be noted that the systematic error in yield monitoring of the $^{100}\rm{Mo}(\gamma,n)^{99}\rm{Mo}$ reaction stems from three unavoidable errors, each running to $\sim$ 1\%. These are the unidentified isotopic composition of natural molybdenum, the uncertainty in the $\gamma$-line intensity used, $I_{\gamma}$ \cite{14}, and the statistical error in determination of the area under the normalizing $\gamma$-line peak. In our calculations we have used the percentage value of $^{100}\rm{Mo}$ isotope abundance equal to 9.63\% (see ref. \cite{12}). 

So, the experimental error of the obtained data ranges between 6\% and 8\%, except for the low-energy case, where the error may amount to $\sim$ 17--20\%.  
     
      \section{Calculation of average cross-section values for the $^{27}\!\rm{Al}(\gamma,\textit{x})^{22}\rm{Na}$ reaction}
      \label{Calc of ave}   
  
     The total and partial cross-sections  $\sigma(E)$ of the $^{27}\!\rm{Al}(\gamma,\textit{x})^{22}\rm{Na}$ reaction were computed  for the monochromatic photons with the TALYS1.95 code \cite{10}, set in Linux Ubuntu-20.04. The computations were performed for different level density ($LD$) models. The TALYS1.95 includes 3 phenomenological level density models and 3 options for microscopic level densities:

$LD 1$: Constant temperature + Fermi gas model;

$LD 2$: Back-shifted Fermi gas model; 

$LD 3$: Generalized superfluid model;

$LD 4$: Microscopic level densities (Skyrme force) from Goriely’s tables; 

$LD 5$: Microscopic level densities (Skyrme force) from Hilaire’s combinatorial tables;  

$LD 6$: Microscopic level densities (temperature dependent HFB, Gogny force) from Hilaire’s combinatorial tables. 

     The TALYS1.95-computed cross-sections $\sigma(E)$ were then averaged over the bremsstrahlung flux $W(E,E_{\rm{\gamma max}})$ in the energy range from the threshold energy of a certain reaction channel, $E_{\rm{th}}$, up to the maximum energy of the bremsstrahlung gamma spectrum, $E_{\rm{\gamma max}} = $ 35--95~MeV. Thus, the flux-averaged cross-section values were computed by the following expression:
           
\begin{equation}\label{form1}
\langle{\sigma(E_{\rm{\gamma max}})}\rangle = \frac
{\int\limits_{E_{\rm{th}}}^{E_{\rm{\gamma max}}}\sigma(E)\cdot W(E,E_{\rm{\gamma max}})dE}
{\int\limits_{E_{\rm{th}}}^{E_{\rm{\gamma max}}}W(E,E_{\rm{\gamma max}})dE}.
\end{equation}

The $\langle{\sigma(E_{\rm{\gamma max}})}\rangle$ values, calculated in this way, were compared with the experimental average cross-sections, which were determined from the expression:
\begin{equation}\begin{split}
\langle{\sigma(E_{\rm{\gamma max}})}\rangle = \; \; \; \; \; \; \; \; \; \; \; \; \; \; \; \; \; \; \; \; \; \; \; \; \; \; \; \; \; \; \; \;\; \; \; \; \; \; \; \; \; \; \; \; \; \; \; \;\; \; \; \; \; \; \; \; \; \; \; \; \; \; \; \; \; \; \; \; \; \;   \\
\frac{\lambda \triangle A}{N_x I_{\gamma} \ \varepsilon \Phi(E_{\rm{\gamma max}}) (1-\exp(-\lambda t_{\rm{irr}}))\exp(-\lambda t_{\rm{cool}})(1-\exp(-\lambda t_{\rm{meas}}))},
\label{form2}
\end{split}\end{equation}
where $\triangle A$ is the number of counts of $\gamma$-quanta in the full absorption peak (for the $\gamma$-line of the investigated reaction), $ {\rm{\Phi}}(E_{\rm{\gamma max}}) = {\int\limits_{E_{\rm{th}}}^{E_{\rm{\gamma max}}}W(E,E_{\rm{\gamma max}})dE}$  is the sum of bremsstrahlung quanta in the energy range from the reaction threshold $E_{\rm{th}}$ up to $E_{\rm{\gamma max}}$, $N_x$ is the number of target atoms, $I_{\gamma}$ is the absolute intensity of the analyzed $\gamma$-quanta, $\varepsilon$ is the absolute detection efficiency for the analyzed $\gamma$-quanta energy, $\lambda$ is the decay constant \mbox{($\rm{ln}2/\textit{T}_{1/2}$)}; $t_{\rm{irr}}$, $t_{\rm{cool}}$ and $t_{\rm{meas}}$ are the irradiation time, cooling time and measurement time, respectively. It is obvious from eqs. \ref{form1} and \ref{form2} that the average cross-section $\langle{\sigma(E_{\rm{\gamma max}})}\rangle$ value is dependent on both the bremsstrahlung flux energy distribution and the reaction threshold energy $E_{\rm{th}}$. 

\begin{figure}[h]
	\center{\includegraphics[scale=0.30]{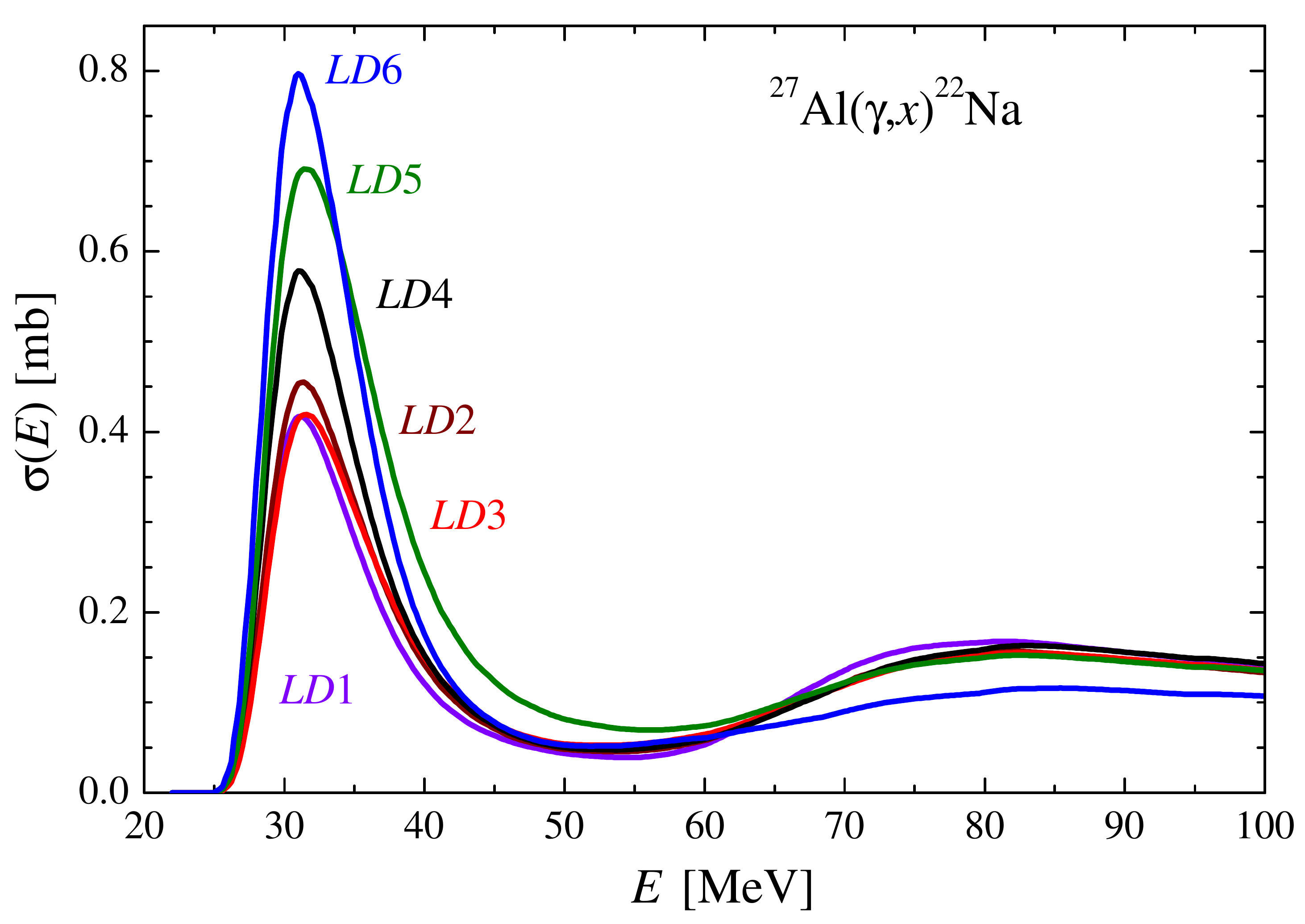}}
	\caption{Total cross-sections $\sigma(E)$  for $^{22}\rm{Na}$ production in the $^{27}\!\rm{Al}(\gamma,\textit{x})^{22}\rm{Na}$ reaction. Computations were made in the TALYS1.95 code with different level density models $LD$ 1-6.}
	\label{fig2a}
\end{figure}

     The TALYS1.95 computation data on the total cross-sections with different level density models $LD$ 1-6 are presented in Fig.~\ref{fig2a}. It can be seen from the figure that the difference between the reaction cross-sections computed by different models reaches a factor of 2 in the vicinity of the cross-section maximum ($\sim$ 32 MeV). Note that the cross-section values at photon energies higher than 50 MeV differ insignificantly, except the $LD$6 case, where the computed value is appreciably lower ($\sim$ 30\%). Thus, in the $LD$ 1-6 models, the relationship among different partial channels of the $^{27}\!\rm{Al}(\gamma,\textit{x})^{22}\rm{Na}$ reaction is very much different. 

The computations for different level densities are distinguished by the contribution of the dominant $(\gamma,\rm n\alpha)$ channel to the total reaction cross-section, which manifests itself in the energy region between 25 and 45 MeV. With variation of the level density model from $LD$1 to $LD$6 the part of this contribution increases. The $^{27}\!\rm{Al}(\gamma,\textit{x})^{22}\rm{Na}$ total cross-sections values are close for $LD$ 1,2,3, but are noticeably different for $LD$ 4,5,6. Hereafter, to avoid overloading the figures, we show only the $LD$1 values instead of the computations by the models $LD$ 1,2,3. 

\begin{figure}[h]
	\center{\includegraphics[scale=0.32]{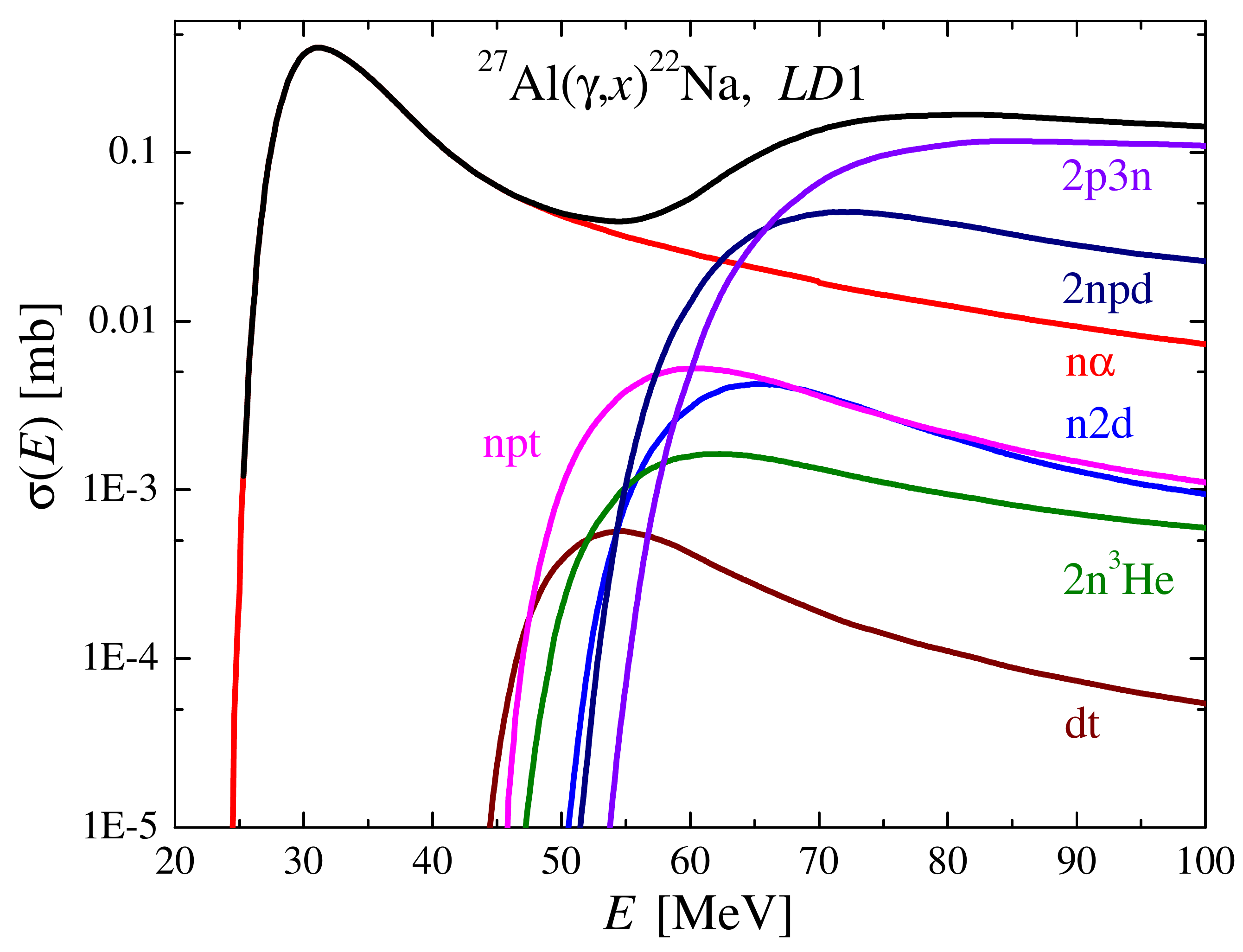}}
	\caption{Cross-sections $\sigma(E)$ for $^{22}\rm{Na}$ production in the  $^{27}\!\rm{Al}(\gamma,\textit{x})^{22}\rm{Na}$ reaction, computed with the TALYS1.95 code for the $LD 1$ model: total cross-section is black curve, and partial cross-section are colored ones.}
	\label{fig3}
\end{figure}

\begin{figure}[h]
	\center{\includegraphics[scale=0.32]{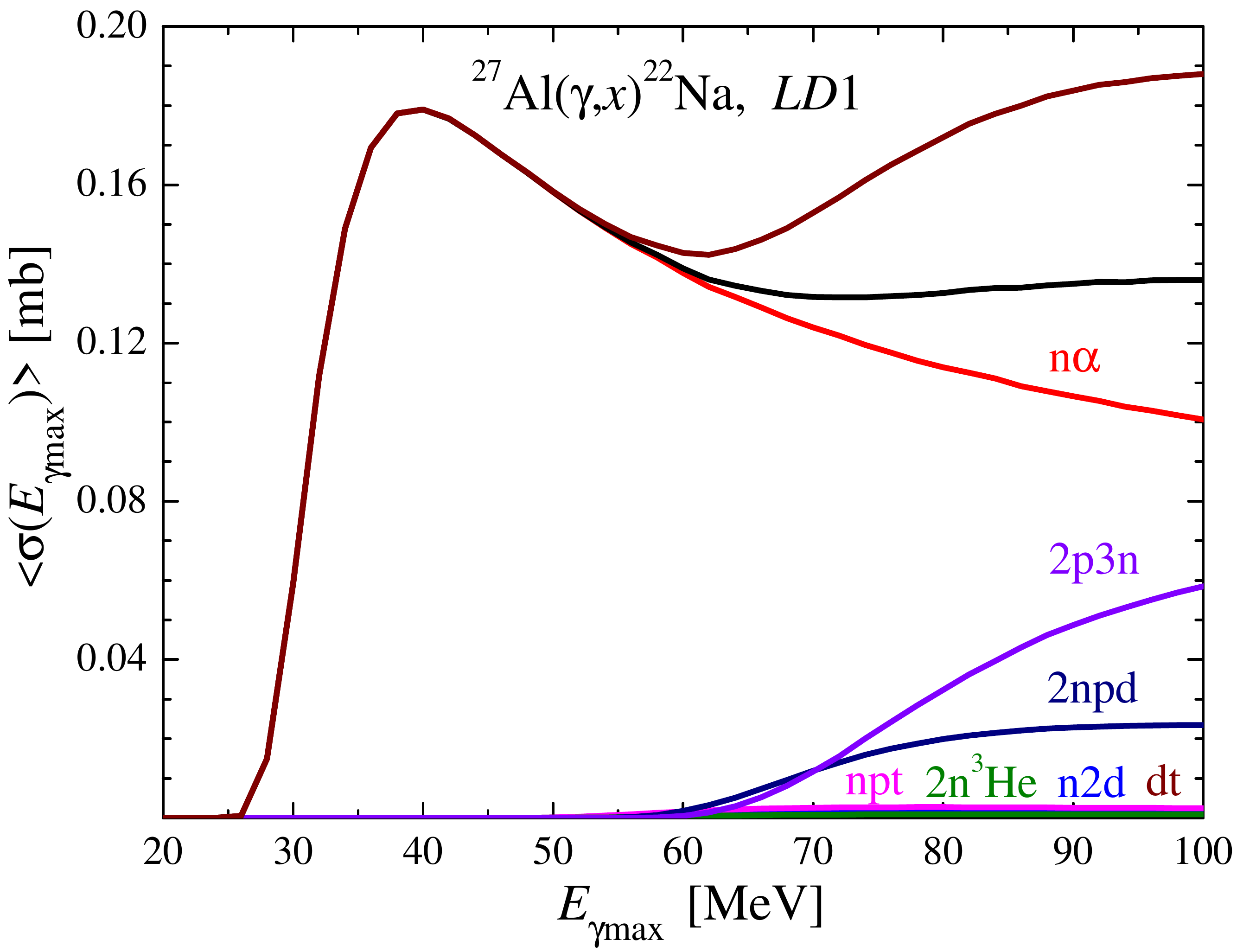}}
	\caption{Flux-averaged cross-sections $\langle{\sigma(E_{\rm{\gamma max}})}\rangle$ for $LD 1$. The total average cross-section is shown in two calculation variants: as a sum of average partial cross-sections with its own $E_{\rm{th}}$ (brown curve), and the total bremsstrahlung flux-averaged cross-section $\sigma(E)$ with $E_{\rm{th}}=$ 22.51 MeV (black curve).}
	\label{fig4}
\end{figure}

By way of example, Fig.~\ref{fig3} shows the total and partial cross-sections $\sigma(E)$ for the $^{27}\!\rm{Al}(\gamma,\textit{x})^{22}\rm{Na}$ reaction, obtained in the TALYS1.95 code with the $LD$1 model. The same cross-sections, but averaged over the bremsstrahlung gamma flux, are shown in Fig.~\ref{fig4}. Note that for calculation of the total flux-averaged cross-section of the reaction under study, it is necessary to add up the average partial cross-sections, each of which is calculated by eq.~(\ref{form1}) with its own threshold $E_{\rm{th}}$.  


The calculation data presented in Figs. \ref{fig2a} to \ref{fig4} show that for both $\sigma(E)$ and flux-averaged $\langle{\sigma(E_{\rm{\gamma max}})}\rangle$ cross-sections, at energies ranging up to 60~MeV, it is the $(\gamma,\rm n\alpha)$ reaction channel that is dominant, while at energies above 70 MeV, the contribution of the $(\gamma,\rm2p3n)$ reaction cross-section becomes noticeable.  In both cases of the calculated cross-sections, $\sigma(E)$ and $\langle{\sigma(E_{\rm{\gamma max}})}\rangle$, the contribution of the $(\gamma,\rm2p3n)$ channel substantially changes the behavior of the energy dependences. As a result, the characteristic minimum is formed in the total cross-section in the energy range between 50 and 60 MeV. 

We now draw your attention to two different variants of calculation of the $^{27}\!\rm{Al}(\gamma,\textit{x})^{22}\rm{Na}$ total average cross section $\langle{\sigma(E_{\rm{\gamma max}})}\rangle$, shown in Fig.~\ref{fig4}. One of them (brown curve) was performed as a sum of partial average cross sections, each being calculated with its own reaction threshold $E_{\rm{th}}$. The black curve shows the total average cross-section $\langle{\sigma(E_{\rm{\gamma max}})}\rangle$ obtained through the use of the bremsstrahlung flux-averaged total cross section $\sigma(E)$, taken from the minimal reaction threshold $E_{\rm{th}}$ = 22.51 MeV. This calculation is in principle incorrect, because eq.~(\ref{form1}) calls for substitution of different reaction thresholds $E_{\rm{th}}$ for each partial cross-section. As is clear from Fig.~\ref{fig4}, there is considerable difference between the two total average cross-sections $\langle{\sigma(E_{\rm{\gamma max}})}\rangle$ after 60~MeV.

Experimentally, for average cross-section determination, the flux value calculated from $E_{\rm{th}}$ up to $E_{\rm{\gamma max}}$ (eq.~(\ref{form2})) is used. In case of a few reaction thresholds, several different fluxes will arise, which must be taken into account in eq.~(\ref{form2}). To calculate the correct $\langle{\sigma(E_{\rm{\gamma max}})}\rangle$ value, a special correction factor ($RK(E_{\rm{\gamma max}})$) should be introduced. This factor is defined as the ratio of two averaged cross-sections (see brown and black curves in Fig.~\ref{fig4}): one calculated as a sum of average partial cross-sections with proper thresholds  $E_{\rm{th}}$ for each reaction channel, and the other -- the total cross-section $\sigma(E)$ averaged for $E_{\rm{th}}$ = 22.51 MeV. Note that for each model of $LD$ 1-6, the factor $RK(E_{\rm{\gamma max}})$ values differ, since the ratios of contributions from different reaction channels are not the same. 

It is evident from Fig.~\ref{fig4} that the correct calculation of the total average cross-section results in its additional increase at the energy beyond 65 MeV, this being due to the reduction in the gamma-flux value in the numerator of eq.~(\ref{form1}) for high threshold reactions.

It is obvious that the dependence of the $\langle{\sigma(E_{\rm{\gamma max}})}\rangle$ value on the reaction threshold complicates the procedure of experimental cross-section determination. This situation occurs only for multiparticle reactions with charged particle yields. In the case of photoneutron reactions, the threshold is uniquely determined, and the average cross-sections are calculated correctly.

For representing the experimental photonuclear reaction data, the average cross-section per equivalent photon is also used. It is determined, similarly to ref.~\cite{17}, by the expression 

\begin{equation}\label{form3}
\langle{\sigma(E_{\rm{\gamma max}})}_{\rm{Q}}\rangle = E_{\rm{\gamma max}}\frac
{\int\limits_{0}^{E_{\rm{\gamma max}}}\sigma(E) \cdot W(E,E_{\rm{\gamma max}})dE}
{\int\limits_{0}^{E_{\rm{\gamma max}}}E\cdot W(E,E_{\rm{\gamma max}})dE}.
\end{equation}

The comparison between two types of the average cross-sections shows the advantage of using $\langle{\sigma(E_{\rm{\gamma max}})_{\rm{Q}}}\rangle$ in the case, when at a certain  $E_{\rm{\gamma max}}$ value several reaction channels with different thresholds $E_{\rm{th}}$ appear. The calculation of this experimental value needs no correction. At that, to level out the influence of different thresholds, the experimental average cross-sections $\langle{\sigma(E_{\rm{\gamma max}})}\rangle$ must be corrected with the use of a specially calculated  $RK(E_{\rm{\gamma max}})$ factor.

  \section{Results and discussion}
 \subsection{Bremsstrahlung flux-averaged cross-sections $\langle{\sigma(E_{\rm{\gamma max}})}\rangle$ for the  $^{27}\!\rm{Al}(\gamma,\textit{x})^{22}\rm{Na}$ reaction}
  \label{Res_disc 1}

\begin{figure}
	\center{\includegraphics[scale=0.3]{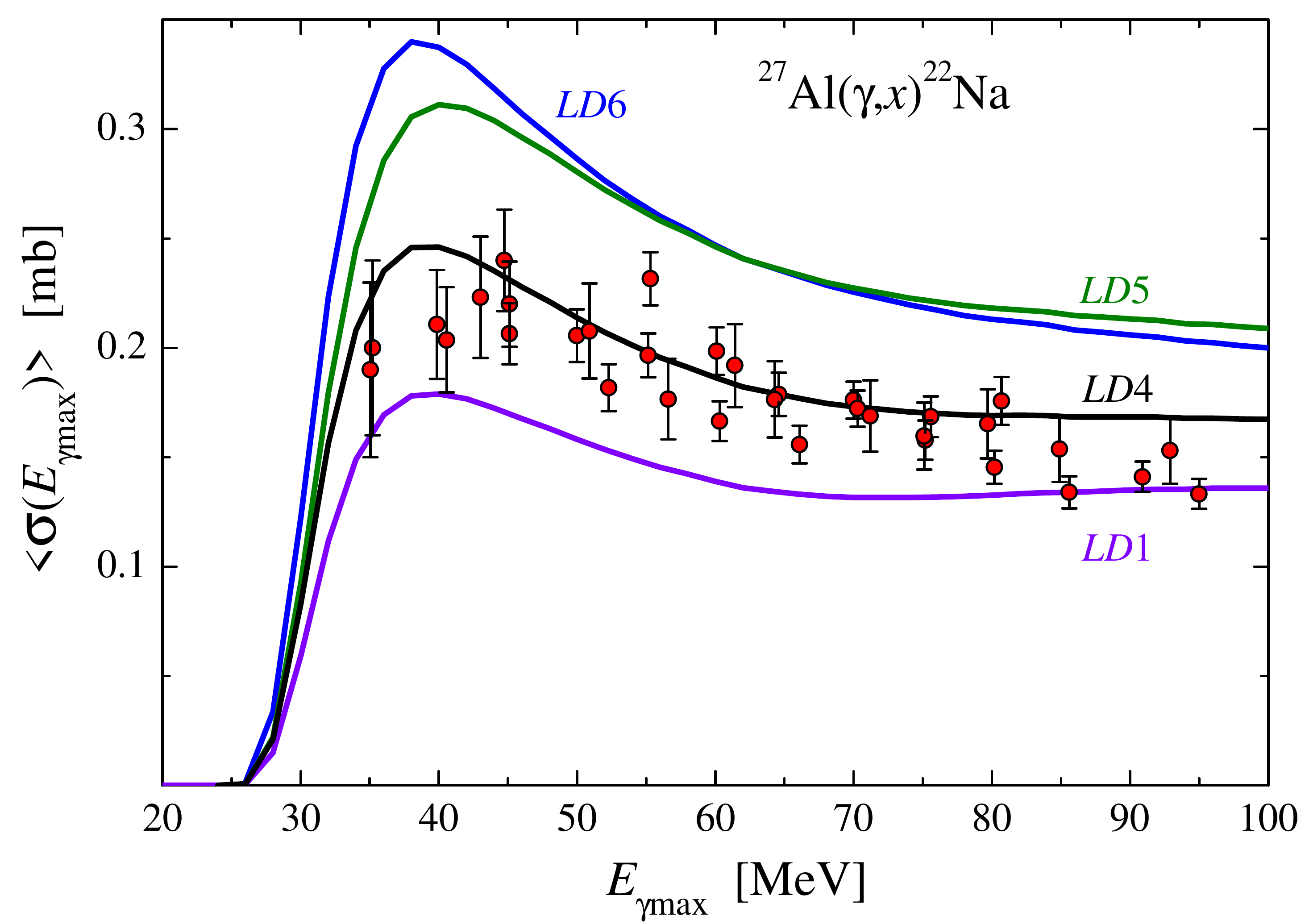}}
	\caption{Total average cross-sections $\langle{\sigma(E_{\rm{\gamma max}})}\rangle$ for the reaction $^{27}\!\rm{Al}(\gamma,\textit{x})^{22}\rm{Na}$, calculated with the threshold $E_{\rm th}$ = 22.51 MeV: red points  --  our data; the curves represent the computations by the TALYS1.95 code with the models \textit{LD} 1, 4, 5, 6.}
	\label{fig6}
	\vspace{2ex}
\end{figure}

Figure~\ref{fig6} shows the data on the total average cross section $\langle{\sigma(E_{\rm{\gamma max}})}\rangle$ for the $^{27}\!\rm{Al}(\gamma,\textit{x})^{22}\rm{Na}$ reaction, which were obtained both experimentally and computationally (with the TALYS1.95 code). For the $\langle{\sigma(E_{\rm{\gamma max}})}\rangle$ calculations by eqs.~(\ref{form1}) and (\ref{form2}), we have chosen the threshold energy of the dominant $^{27}\!\rm{Al}(\gamma,\rm{n}\alpha)^{22}\rm{Na}$ channel, $E_{\rm{th}}$ = 22.51 MeV. 

The experimental $\langle{\sigma(E_{\rm{\gamma max}})}\rangle$ values at energies ranging from 45 MeV to 80 MeV are in satisfactory agreement with the values computed by the version with the $LD4$ model, while at energies of 35 to 40 MeV and above 80 MeV, the obtained experimental data lie between the computed $LD1$ and $LD4$ curves.  The values calculated with the $LD 5$ and $LD 6$ models are systematically found above the experimental data.

\begin{figure}
	\center{\includegraphics[scale=0.3]{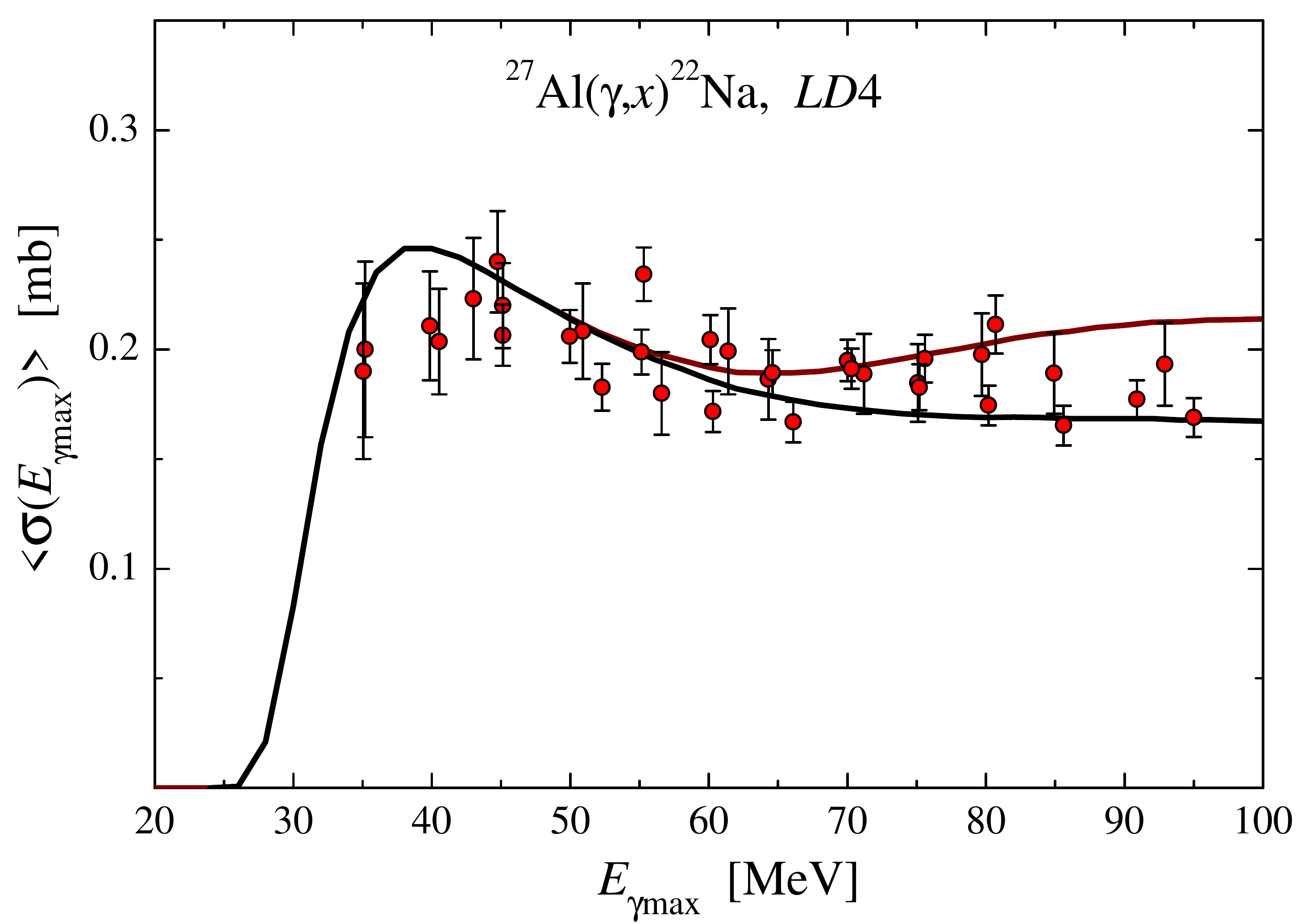}}
	\caption{Total average cross-sections $\langle{\sigma(E_{\rm{\gamma max}})}\rangle$ for the reaction $^{27}\!\rm{Al}(\gamma,\textit{x})^{22}\rm{Na}$. The TALYS1.95 (\textit{LD}4) computations were performed for $E_{\rm th}$ = 22.51 MeV (black curve), and as a sum of partial average cross-sections (brown curve). The points represent the experimental data corrected for the   $RK(E_{\rm{\gamma max}})$ factor for the level density model  \textit{LD}4.}
	\label{fig7}
\end{figure}

Figure ~\ref{fig7} shows the total average cross-sections determined with the use of two calculation variants: one, calculated as a sum of average partial cross-sections (brown curve); and the other -- the bremsstrahlung flux-averaged total cross-section $\sigma(E)$ with $E_{\rm{th}}$ = 22.51~MeV (black curve). As can be seen, the $RK(E_{\rm{\gamma max}})$ factor is equal to 1 at low energies, but beginning with 55~MeV, it smoothly increases with the energy $E_{\rm{\gamma max}}$, reaching 1.27 at 95~MeV for the $LD 4$ model. As mentioned above, the $RK(E_{\rm{\gamma max}})$ value varies depending on the level density model. Its growth rate is maximal for the model $LD 1$ (reaching 1.35 at 95~MeV), falls off with an increase in the model number, and equals 1.16 for the $LD 6$ model (at 95~MeV).

The experimental data were multiplied by the $RK(E_{\rm{\gamma max}})$ factor for the calculation variant $LD 4$, and are given in Fig.~\ref{fig7}. As is seen from the figure, the experimental $\langle{\sigma(E_{\rm{\gamma max}})}\rangle$ values lie close to the brown curve. The experimental points at energies lower than 40~MeV and higher than 80~MeV are found somewhat below the mentioned curve. It must be emphasized that the correct experimental data are the ones obtained with the application of the $RK(E_{\rm{\gamma max}})$ factor, and shown in Fig.~\ref{fig7}. 

The analysis of Figs.~\ref{fig6} and \ref{fig7} suggests the conclusion that within the measurement accuracy the experimental data lie most closely to the calculation using the cross-sections $\sigma(E)$ determined with the TALYS1.95 code for the $LD 4$ model.

 \subsection{Cross-sections per equivalent photon $\langle{\sigma(E_{\rm{\gamma max}})_{\rm{Q}}}\rangle$ for the $^{27}\!\rm{Al}(\gamma,\textit{x})^{22}\rm{Na}$ reaction}
  \label{Res_disc 2}

The experimental data available in the literature on the $^{27}\!\rm{Al}(\gamma,\textit{x})^{22}\rm{Na}$ reaction were obtained in terms of the average cross-section per equivalent photon, refs.~\cite{8,9}. For comparison with those data, we have also represented our present results in the form of $\langle{\sigma(E_{\rm{\gamma max}})_{\rm{Q}}}\rangle$ in accordance to eq.~(\ref{form3}). 

\begin{figure}[h]
	\center{\includegraphics[scale=0.3]{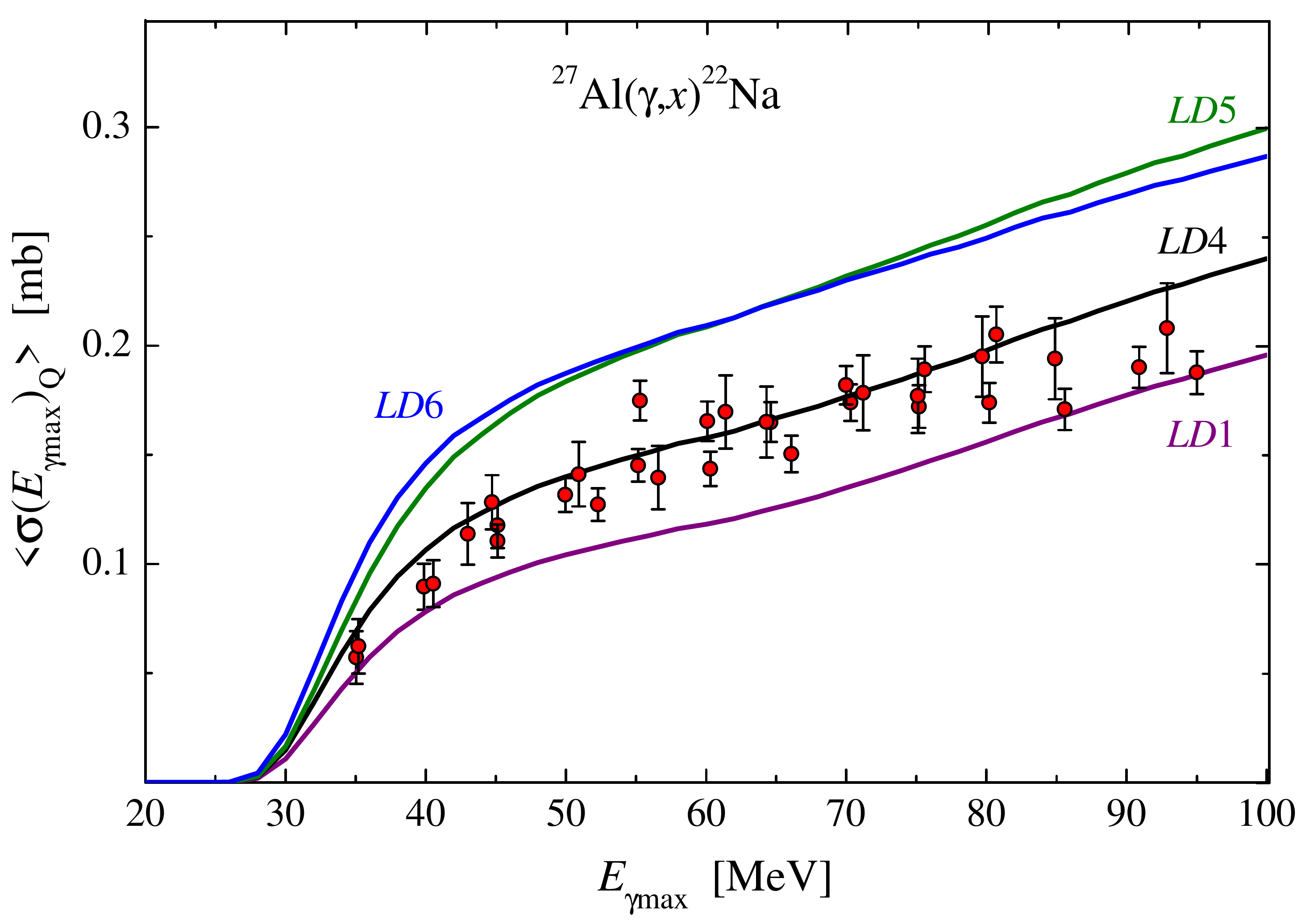}}
	\caption{Total cross-sections per equivalent photon $\langle{\sigma(E_{\rm{\gamma max}})_{\rm{Q}}}\rangle$ for the $^{27}\!\rm{Al}(\gamma,\textit{x})^{22}\rm{Na}$ reaction. Red points – experimental values; the curves show the TALYS1.95 computations for \textit{LD} 1, 4, 5, 6 models.}
	\label{fig8}
\end{figure}

Figure \ref{fig8} shows the experimental values of the total average cross-section $\langle{\sigma(E_{\rm{\gamma max}})_{\rm{Q}}}\rangle$ for the  $^{27}\!\rm{Al}(\gamma,\textit{x})^{22}\rm{Na}$ reaction, and also, the data computed with the use of the codes TALYS1.95 and GEANT4.9.2. As in the case of the average cross-section $\langle{\sigma(E_{\rm{\gamma max}})}\rangle$, the experimental $\langle{\sigma(E_{\rm{\gamma max}})_{\rm{Q}}}\rangle$ values lie between the calculated curves for the  $LD 1$ and 4 models.

\begin{figure}[h]
	\center{\includegraphics[scale=0.3]{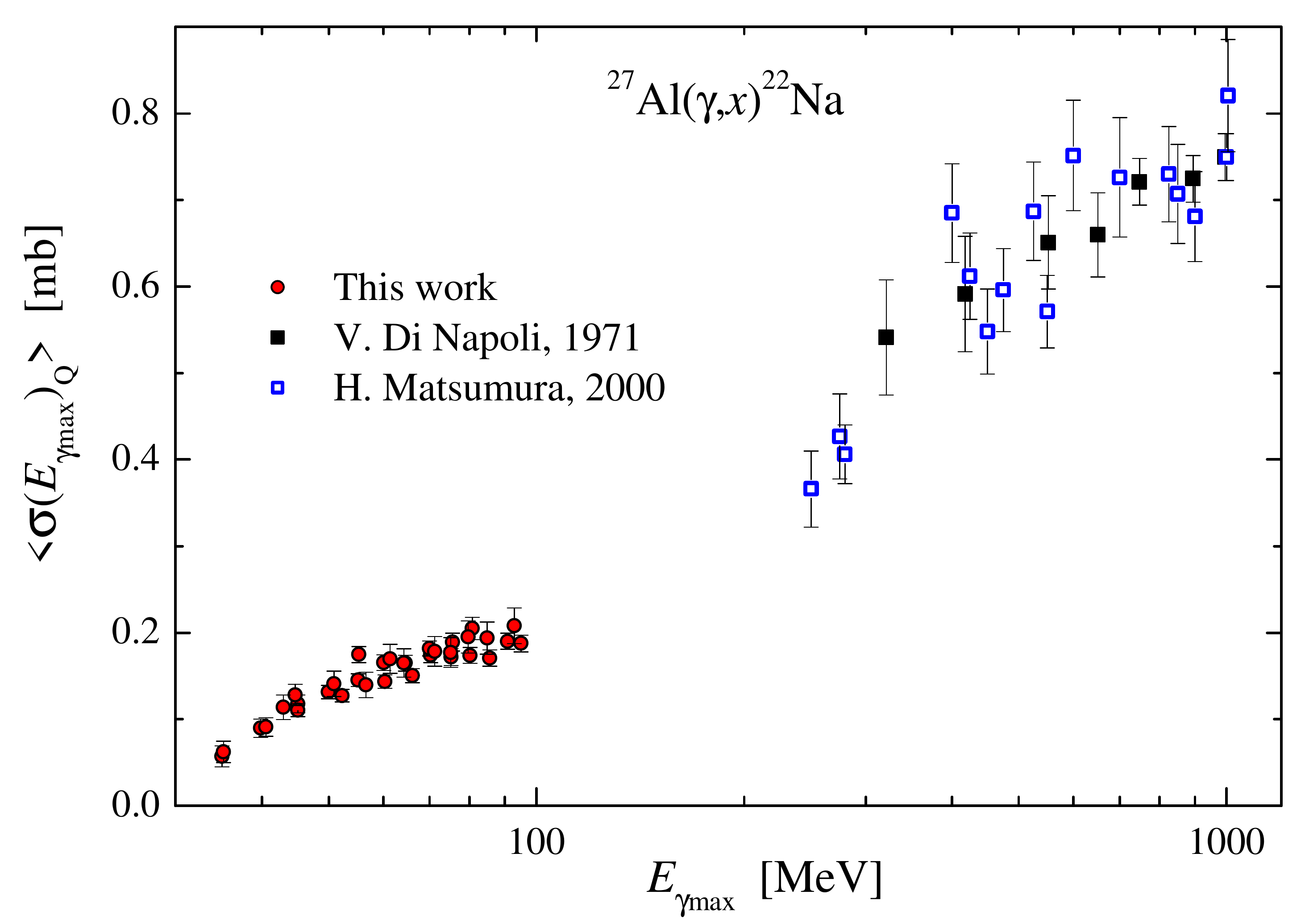}}
	\caption{Experimental values of total cross-sections per equivalent photon $\langle{\sigma(E_{\rm{\gamma max}})_{\rm{Q}}}\rangle$ for the $^{27}\!\rm{Al}(\gamma,\textit{x})^{22}\rm{Na}$  reaction: red points  -- present work, full squares -- ref. \cite{8}, empty squares -- ref. \cite{9}.}
	\label{fig9}
\end{figure}

Figure \ref{fig9} shows the experimental $\langle{\sigma(E_{\rm{\gamma max}})_{\rm{Q}}}\rangle$ data obtained in refs.~\cite{8,9} at gamma-ray energies ranging from 250 to 1000~MeV.  Our present results are an addition to the data of \cite{8,9} for the 35--95~MeV range.  

All experimental data available for the reaction $^{27}\!\rm{Al}(\gamma,\textit{x})^{22}\rm{Na}$ show that on a semilogarithmic scale the behavior of $\langle{\sigma(E_{\rm{\gamma max}})_{\rm{Q}}}\rangle$ as a function of energy exhibits a more complex structure  as opposed to the linear dependence, previously assumed in ref.~\cite{8} for the energy range from 320 to 1000~MeV. This is due to the change in the mechanism of gamma-quanta interaction with the nucleus at different energy intervals. Note that for the multiparticle photonuclear reaction 
$^{27}\!\rm{Al}(\gamma,\textit{x};\textit{x}= {^3\rm{He}} + pd + 2pn)^{24}\rm{Na}$
 the linear energy dependence of $\langle{\sigma(E_{\rm{\gamma max}})_{\rm{Q}}}\rangle$ had also been presumed within the range 300 to 1000~MeV, however the analysis for a wider energy interval has pointed to its more complicated character \cite{4,8}.

\section{Conclusions}
\label{Concl}

The present work has been concerned with the first measurements of the total average cross-sections $\langle{\sigma(E_{\rm{\gamma max}})}\rangle$, and the cross-sections per equivalent photon $\langle{\sigma(E_{\rm{\gamma max}})_{\rm{Q}}}\rangle$ for the multichannel reaction $^{27}\!\rm{Al}(\gamma,\textit{x};\textit{x}=\rm{n}\alpha + dt + npt + 2n{^{3}\rm{He}} + 2nd + 2npd + 2p3n)^{22}\rm{Na}$ in the energy range $E_{\rm{\gamma max}}$ = 35--95~MeV with the use of the induced $\gamma$-activity method. The bremsstrahlung gamma-quantum flux was computed in the GEANT4.9.2 code, and in addition, was monitored by means of the $^{100}\rm{Mo}(\gamma,n)^{99}\rm{Mo}$ reaction yield. The obtained $\langle{\sigma(E_{\rm{\gamma max}})_{\rm{Q}}}\rangle$ values are compared with the data of other laboratories. 

The partial cross-sections $\sigma(E)$ for the reaction under study were calculated using the TALYS1.95 code for 6 variants of the level density model (3 phenomenological level density models and 3 options for microscopic level densities) -- $LD$ 1--6. 

It has been demonstrated that at energies up to 60~MeV the dominant contribution comes from $^{27}\!\rm{Al}(\gamma,n\alpha)^{22}\rm{Na}$, whereas at 70--95~MeV, the $^{27}\!\rm{Al}(\gamma,2p3n)^{22}\rm{Na}$ reaction also contributes essentially. The threshold difference between these reactions amounts to 28.3~MeV. This allows one to investigate the relative contribution of different $^{27}\!\rm{Al}(\gamma,\textit{x})^{22}\rm{Na}$ reaction channels over different energy ranges. 
 
In the case of the multiparticle reaction with charged particles in the outlet channel, such as the $^{27}\!\rm{Al}(\gamma,\textit{x})^{22}\rm{Na}$ reaction, the correct calculation of the total average cross-sections $\langle{\sigma(E_{\rm{\gamma max}})}\rangle$ is provided through their computation with the TALYS1.95 code as a sum of partial average cross-sections, each having been calculated with its own reaction threshold $E_{\rm th}$. The experimental $\langle{\sigma(E_{\rm{\gamma max}})}\rangle$ values are calculated through averaging the cross-sections over the bremsstrahlung flux taken from the minimum threshold $E_{\rm th}$ = 22.51~MeV, and then correcting the obtained values for the corresponding $RK(E_{\rm{\gamma max}})$ factor. The $RK(E_{\rm{\gamma max}})$ value was defined as the ratio of two averaged cross-sections: \textit{one}, calculated as a sum of average partial cross-sections with proper thresholds  $E_{\rm th}$ for each reaction channel, and \textit{the other}, the total cross-section $\sigma(E)$ averaged for $E_{\rm th}$ = 22.51 MeV. 

The use of the $\langle{\sigma(E_{\rm{\gamma max}})}\rangle$ value permits a more detailed consideration of the bremsstrahlung gamma energy dependence of the cross-section, since it is insensitive to the low-energy part of the bremsstrahlung spectrum. On the other hand, the cross-sections per equivalent quantum $\langle{\sigma(E_{\rm{\gamma max}})_{\rm{Q}}}\rangle$ are proportional to the reaction yield and always increase with the increasing end-point bremsstrahlung energy $E_{\rm{\gamma max}}$. The chief flaw of the representation of cross-sections per equivalent quantum $\langle{\sigma(E_{\rm{\gamma max}})_{\rm{Q}}}\rangle$ is the influence of bremsstrahlung quanta having energies below the threshold of the reaction under study. 

Thus, the experimental values of total average cross-sections $\langle{\sigma(E_{\rm{\gamma max}})}\rangle$ and $\langle{\sigma(E_{\rm{\gamma max}})_{\rm{Q}}}\rangle$ have been shown to be in satisfactory agreement with the calculation by the $LD 4$ model. The results of the present work may be considered as a useful addition to the data available in the literature for the energy range 35 to 95~MeV. 

\section*{Acknowledgment}
The authors would like to thank the staff of the linear electron accelerator LUE-40 NSC KIPT, Kharkiv, Ukraine, for their cooperation in the realization of the experiment.

\section*{Declaration of competing interest}\addcontentsline{toc}{section}{Declaration of competing interest}
\label{Decl}
The authors declare that they have no known competing financial interests or personal relationships that could have appeared to influence the work reported in this paper.



\end{document}